\documentclass[12pt,twoside,a4paper]{article} 
\usepackage{amsmath,amssymb,latexsym,theorem,bbm,shapepar,natbib,epsfig}
\newfont{\msa}{msam10 scaled\magstep1}
\newfont{\ssmsa}{msam9}

\def\crps{\mathop{\hbox{\rm crps}}}

\numberwithin{equation}{section}

\title{Comparison of BMA and EMOS statistical calibration methods for
temperature and wind speed ensemble weather prediction}

\author{{\sc S\'andor Baran$^{1}$}, {\sc Andr\'as Hor\'anyi$^{2}$} \\
  and {\sc D\'ora Nemoda$^{1}$}\\ 
         $^1$Faculty of Informatics, University of Debrecen\\
         Kassai \'ut 26, H-4028 Debrecen, Hungary \\                
         $^2$ Hungarian Meteorological Service \\
       P.O. Box 38, H--1525 Budapest, Hungary 
     }

\date{}

\begin{document}

\pagestyle{myheadings}

\maketitle

\begin{abstract}
The evolution of the weather can be described by deterministic numerical weather
forecasting models. Multiple runs of these models with different
initial conditions and/or model physics result in forecast ensembles
which are used for estimating the distribution of future atmospheric
variables. However, these ensembles are usually under-dispersive and
uncalibrated, so post-processing is required. 

In the present work we compare different versions of Bayesian Model Averaging
(BMA) and Ensemble Model Output Statistics (EMOS) post-processing methods in
order to calibrate 2m temperature and 10m wind speed forecasts of the
operational ALADIN Limited Area Model Ensemble Prediction System
of the Hungarian Meteorological Service.
We show that compared to the raw ensemble both post-processing methods improve
the calibration of probabilistic and accuracy of point forecasts and
that the best BMA method slightly outperforms the EMOS technique.

\end{abstract}

\section{Introduction}
   \label{sec:sec1} 
The evolution of the weather can be described by Numerical Weather
Prediction (NWP) models, which are capable to simulate the atmospheric
motions taking into account the physical governing laws of the
atmosphere and the connected spheres (typically sea or land
surface). Without any doubts these models provide primary support for
meteorological forecasting and decision making. As a matter of fact the NWP
models and consequently the weather forecasts cannot be fully precise
and on top of that their precision might change with the
meteorological situation as well (due to the chaotic character of the
atmosphere, which manifests in being very sensitive to its initial
conditions). Therefore, it is a relevant request from the users to
provide uncertainty estimations attached to the weather forecasts. The
information related to the intrinsic uncertainty of the weather
situation and the model itself is very valuable additional
information, which is generally provided by the use of ensemble
technique. The ensemble method is based on the accounting of all
uncertainties exist in the NWP modelling process and its expression in
terms of forecast probabilities. These probabilities are provided by
the various realizations of the weather forecasting models, which are
differently modified based on the underlying uncertainties. From the
practical point of view the ensemble technique exploits several NWP
model runs (and these ensemble model members differ within the known
uncertainties of the initial and boundary conditions, model
formulation etc.) and then evaluates the ensemble of forecasts
statistically. One possible improvement area of the ensemble forecasts
is the calibration of the ensemble in order to transform the original
ensemble member-based probability density function (PDF) into a more
reliable and realistic one. The main disadvantage of the method is
that it is based on statistics of model outputs, and therefore unable
to consider the physical aspects of the underlying processes. The
latter issues should be addressed by the improvements of the reality
of the model descriptions and particularly the better uncertainty
descriptions used by the different model realizations.

From the various modern post-processing techniques probably the most
popular methods are the Bayesian model averaging 
\citep[BMA, see e.g.][]{rgbp,srgf,sgr10} and the 
ensemble model output statistics \citep[EMOS, see e.g.][]{grwg,wh,tg} which are 
implemented in {\tt ensembleBMA} \citep{frgs,frgsb} and {\tt ensembleMOS}
packages of {\tt R}. Both approaches provide estimates of the densities
of the predictable weather quantities and once a predictive density is
given, a point forecast can be easily determined (e.g. mean or median value). 

The BMA method for calibrating 
forecast ensembles was introduced by \citet{rgbp}. The BMA predictive
PDF of a future weather quantity is the
weighted sum of individual PDFs corresponding to the ensemble
members. An individual PDF can be interpreted as the
conditional PDF of the future weather quantity provided the considered
forecast is the best one and the weights are based on the
relative performance of the ensemble members during a given training
period. 

The EMOS approach, proposed by \citet{grwg}, uses a single parametric
distribution as a predictive PDF with parameters depending on the
ensemble members. 

In both post-processing techniques the unknown parameters are estimated using
forecasts and validating observations from a rolling training period,
which allows automatic adjustments of the statistical model to any
changes of the EPS system (for instance seasonal variations or EPS
model updates). EMOS method 
is usually more parsimonious and computationally more effective than
BMA, but shows less flexibility. E.g. in case of a weather quantity
following normal or truncated 
normal distribution the EMOS predictive PDF is by definition unimodal,
while BMA approach allows multimodality.

The aim of the present paper is to compare the performance of BMA and EMOS
calibration on the ensemble forecasts of temperature and wind speed
produced by the operational 
Limited Area Model Ensemble   
Prediction System (LAMEPS) of the Hungarian Meteorological Service
(HMS) called ALADIN-HUNEPS \citep{hagel, horanyi}.

\section{ALADIN-HUNEPS ensemble}
  \label{sec:sec2}

\begin{figure}[t]
\begin{center}
\leavevmode
\epsfig{file=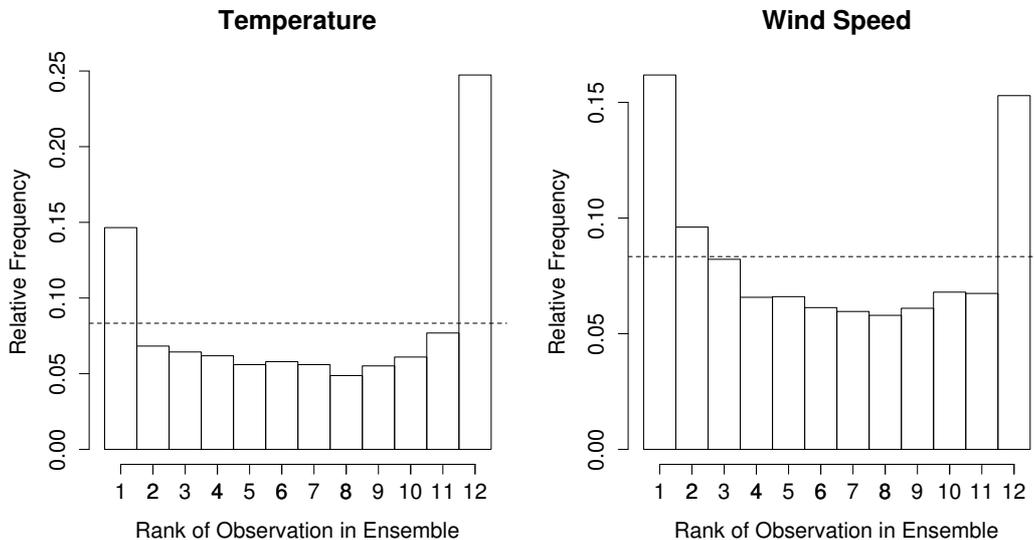,height=7cm, angle=0}
\caption{Verification rank histograms of the 11-member ALADIN-HUNEPS
  ensemble forecasts of 2m temperature and 10m wind speed.  Period: April 1,
  2012 --  March 31, 2013.}  
\label{fig:fig1}
\end{center}
\end{figure}
The ALADIN-HUNEPS system of the HMS covers a large part of Continental
Europe with a horizontal resolution 
of 12 km and it is obtained with dynamical downscaling (by the ALADIN
limited area model) of the global
ARPEGE based PEARP system of M\'et\'eo France \citep{hkkr,dljn}. The
ensemble consists of 11 members, 10 initialized from perturbed initial
conditions and one control member from the unperturbed analysis,
implying that the ensemble contains groups of exchangeable
forecasts. 

The initial perturbations for PEARP are generated with the combination
of singular vector-based and EDA-based perturbations \citep{ldcm}. The singular
vectors are optimized for 7 subdomains and then combined into
perturbations. The EDA perturbations are computed as differences
between the EDA members and the EDA mean (there is a 6-members EDA
system running in France). These two sets of perturbations are
combined into 17 perturbations, which are added to and subtracted from
the control initial condition. Random sets of physical
parameterizations (there are 10 sets of different physical
parameterization packages) are attributed to the forecasts run from
the differently perturbed initial conditions. All these combinations
result in a 35-members (one control without perturbation and 34
perturbed members) global ensemble. The ALADIN-HUNEPS system simply
takes into account (and dynamically downscale) the control and the
first 10 members of the PEARP system. These members contain the first
5 global perturbations added to and subtracted from the control.    

The database contains 11 member ensembles of 42 hour 
forecasts for 2 meter temperature (given in K) and 10 meter wind speed
(given in m/s) for 10 major cities in 
Hungary (Miskolc, Szombathely, Gy\H or, Budapest, Debrecen, Ny\'\i regyh\'aza,
Nagykanizsa, P\'ecs, Kecskem\'et, Szeged) produced by the
ALADIN-HUNEPS system of the HMS, together with the corresponding
validating observations for the one-year period between April 1, 2012
and March 31, 2013. The forecasts are
initialized at 18 UTC. The data set is fairly complete
since there are only six days when no forecasts are available. These dates were
excluded from the analysis.  

Figure \ref{fig:fig1} shows the verification rank histograms of the 
ensemble forecasts of temperature and wind speed. These are the
histograms of ranks of validating 
observations with respect to the corresponding ensemble
forecasts computed from the ranks at all locations and over the whole
  verification period \citep[see e.g.][Section 8.7.2]{wilks}. 
Both histograms are 
far from the desired uniform distribution as in many cases the
ensemble members either underestimate or overestimate the validating
observations. The ensemble ranges contain the observed temperature and
wind speed only in $60.61\%$ and $68.52\% $ of the cases, respectively
(while their nominal values equal $10/12$, i.e $83.33 \%$). 
Hence, both ensembles are under-dispersive
and in this way they are uncalibrated. This supports the need of statistical
post-processing in order to improve the forecasted probability
density functions. 

We remark that BMA calibration of
wind speed \citep{bar,bhn1} and temperature \citep{bhn2} forecasts of
the ALADIN-HUNEPS system have already been investigated by the authors
using smaller data sets covering the period from October 1, 2010 to
March 25, 2011. These investigations showed that significant
improvements can be gained with the use of BMA
post-processing. Nevertheless it is interesting to see what
enhancement can be obtained by BMA with respect to an improved raw EPS
system and particularly in comparison to the EMOS calibration
technique. 

\section{Methods and verification scores}
  \label{sec:sec3}

As it has been mentioned in the Introduction, our study is concentrating on
BMA and EMOS approaches. By \ $f_1,f_2,\ldots ,f_M$ \ 
we denote the ensemble forecast of a certain weather quantity
\ $X$ \ for a given location and time. 
The ensemble members are either distinguishable (we can clearly
identify each member or at least some of them) or indistinguishable
(when the origin of the given member cannot be identified). 
Usually, the distinguishable EPS systems are 
the multi-model, multi-analyses ensemble systems, where each
ensemble member can be identified and tracked.  
This property holds e.g. for the 
University of Washington mesoscale ensemble \citep{em05} or for the
COSMO-DE ensemble of the German Meteorological Service (DWD)
\citep{gtpb}.

However, most of the currently used ensemble
prediction systems incorporate ensembles where at least some
members are statistically indistinguishable. Such ensemble systems are
usually producing initial conditions based on algorithms, which are
able to find the fastest growing perturbations indicating the directions
of the largest uncertainties (for instance singular vector
computations \citep{btmp} or search for breeding vectors \citep{tk}).
In most cases these initial perturbations are  further 
enriched by perturbations simulating model uncertainties as well. It is
typically the case for the 51 
member European Centre for Medium-Range Weather Forecasts
ensemble \citep{lp} or for the PEARP and ALADIN-HUNEPS ensemble \citep{hagel,
  horanyi} described in Section \ref{sec:sec2}. In such cases one
usually has a control member (the one without any perturbation) and
the remaining ensemble members forming one or two exchangeable
groups.  

In what follows, if we have \ $M$ \ ensemble members divided
into \ $m$ \ exchangeable 
groups, where the \ $k$th \ group contains \ $M_k\geq 1$ \ ensemble
members ($\sum_{k=1}^mM_k=M$), \ notation \ $f_{k,\ell}$
\ is used for the  $\ell$th member of the $k$th group.

\subsection{Bayesian Model Averaging}
  \label{subs:subs3.1}

In the BMA model proposed by
\citet{rgbp}, to each ensemble member \ 
$f_k$ \ corresponds a component PDF \ $g_k(x \vert f_k, \theta_k)$, \
where \ $\theta_k$ \ is a parameter to be estimated. The BMA
predictive PDF of \ $X$ \ is
\begin{equation}
  \label{eq:eq3.1}
p(x\vert\, f_1, \ldots ,f_M):=\sum_{k=1}^M\omega _k g_k(x \vert\, f_k; \theta_k),
\end{equation}
where the weight \ $\omega_k$ \ is connected to the relative
performance of the ensemble member \ $f_k$ \ during the training
period.  Obviously, these weights form a probability distribution, that
is \ $\omega_k\geq 0, \ k=1,2, \ldots ,M,$ \ and \ $\sum_{k=1}^M
\omega_k=1$. \ 

For the situation when \ $M$ \ ensemble members are divided into \ $m$ \
exchangeable 
groups \citet{frg} suggested to use model
\begin{equation}
  \label{eq:eq3.2}
p(x\vert f_{1,1}, \ldots ,f_{1,M_1}, \ldots ,  f_{m,1}, \ldots
,f_{m,M_m}):=\sum_{k=1}^m\sum_{\ell=1}^{M_k}\omega _k g_k(x \vert\,
f_{k,\ell}; \theta_k), 
\end{equation}
where ensemble members within a given group have the same weights
and parameters. Since this is the case for the ALADIN-HUNEPS ensemble
(i.e. it consists of groups of exchangeable members), in what
follows we present only the weather variable specific 
versions of  \eqref{eq:eq3.2}.

\subsubsection*{Temperature}
For modelling temperature (and pressure) \citet{rgbp} and \citet{frg}
use normal component PDFs and model \eqref{eq:eq3.2} takes the
form
\begin{equation}
  \label{eq:eq3.3}
p\big(x\big\vert f_{1,1}, \ldots ,f_{1,M_1}, \ldots ,  f_{m,1}, \ldots
,f_{m,M_m}\big):=\sum_{k=1}^m\sum_{\ell=1}^{M_k}\omega _k g\big(x \big\vert 
f_{k,\ell}; \beta_{0,k},\beta_{1,k},\sigma^2\big), 
\end{equation}
where \ $g\big(x\big\vert f;\beta_0,\beta_1,\sigma^2\big )$ \ is
a normal PDF with mean \ $\beta_0+\beta_1f$ \ (linear bias correction)
and variance \
$\sigma^2$. \ Mean parameters \ $\beta_{0,k}$ \ and \ $\beta_{1,k}$ \
are usually estimated with linear regression of the validating 
observation on the corresponding ensemble members, while weights \
$\omega_k$ \ and variance \ $\sigma^2$, \ by maximum likelihood (ML)
method using training data consisting of ensemble members and verifying
observations from the preceding \ $n$ \ days (training period).
For example, taking \ $n=30$ \ the predictive PDF e.g for 12 UTC March 31,
2013 at a given place can be 
obtained from the ensemble forecast for
this particular day, time  and location (initialized at 18 UTC, March 29,
2013) with model parameters estimated from forecasts and
verifying observations for all 10 locations from the 
period February 28 -- March 29, 2013 (30 days, 300 forecast cases).

Another method for estimating model parameters is to
minimize an appropriate verification score (see
Section \ref{subs:subs3.3}) using the same
rolling training data as before. 

As special cases of model \eqref{eq:eq3.3} one can also consider the
situations when only additive bias correction present, that is \
$b_{1,k}=1$, \ and when bias correction is not applied at all, i.e. \
$b_{0,k}=0$ \ and \ $b_{1,k}=1,\ k=1,2, \ldots, m$.

\subsubsection*{Wind speed}
Since wind speed can take only non-negative values, for modelling this
weather quantity a skewed distribution is required.  A
popular candidate is the Weibull distribution \citep[see
e.g.][]{jhmg}, while \citet{sgr10} proposes the BMA model 
\begin{equation}
  \label{eq:eq3.4}
p(x\vert f_{1,1}, \ldots ,f_{1,M_1}, \ldots ,  f_{m,1}, \ldots
,f_{m,M_m}):=\sum_{k=1}^m\sum_{\ell=1}^{M_k}\omega _k h(x \vert 
f_{k,\ell}; b_{0,k},b_{1,k},c_0,c_1), 
\end{equation}
where by \  $h(x \vert f; b_0,b_1,c_0,c_1)$ \ we denote the
PDF of gamma distribution with mean \ $b_0+b_1f$ \ and 
standard deviation \ $c_0+c_1f$. \ Parameters can be estimated in the
same way as before, that is either mean parameters by regression and weights
and standard deviation parameters by ML method or by minimizing a
verification score. We remark that in the {\tt ensembleBMA} package of
R a more parsimonious model is implemented, where the mean parameters
are constant across all ensemble members. In what follows we will use
this simplification, too.

As an alternative, \citet{bar} suggests to model wind speed with a
mixture of truncated normal distributions with a cut-off at zero \
${\mathcal N}^{\, 0}\big(\mu,\sigma^2\big)$, \ where the location \
$\mu$ \ of a component PDF is an affine function of the corresponding
ensemble member. The proposed BMA model is  
\begin{equation}
  \label{eq:eq3.5}
p\big(x\big\vert f_{1,1}, \ldots ,f_{1,M_1}, \ldots ,  f_{m,1}, \ldots
,f_{m,M_m}\big):=\sum_{k=1}^m\sum_{\ell=1}^{M_k}\omega _k q\big(x \big\vert 
f_{k,\ell}; \beta_{0,k},\beta_{1,k},\sigma^2\big), 
\end{equation}
where \ $q\big(x\big\vert f;\beta_0,\beta_1,\sigma^2\big )$ \ is
a truncated normal PDF  with location \
$\beta_0+\beta_1f$ \  
and scale \ $\sigma^2$, \ 
that is 
\begin{equation*}
q\big(x\big\vert f;\beta_0,\beta_1,\sigma^2\big ):=\frac{\frac
  1{\sigma}\varphi\big((x-\beta_0-\beta_1f)/\sigma\big)}{\Phi
  \big((\beta_0+\beta_1f)/\sigma\big)
}, \qquad \text{for \ $x\geq 0$}, \
\end{equation*}
and \ $0$, \ otherwise. Here \ $\varphi$ \ and \ $\Phi$ \ denote the
PDF and the cumulative 
distribution function (CDF) of the
standard normal distribution, respectively. 

For estimating parameters of model \eqref{eq:eq3.5} \citet{bar}
suggests a full ML method, which means that all parameter estimates
are obtained by 
maximizing the likelihood function corresponding to the training data.

\subsection{Ensemble Model Output Statistics}
  \label{subs:subs3.2}

As noted, the EMOS predictive PDF is a single parametric density where
the parameters are functions of the ensemble members.

\subsubsection*{Temperature}
Similarly to the BMA approach, for modelling temperature (and
pressure) normal distribution seems to be a reasonable choice. The
EMOS predictive distribution suggested by \citet{grwg} is
\begin{equation}
   \label{eq:eq3.6}
  {\mathcal N}\big(a_0+a_1f_1+ \ldots +a_Mf_M,b_0+b_1S^2\big) \qquad
  \text{with} \qquad S^2:=\frac 1{M-1}\sum_{k=1}^M\big (f_k-\overline f\big)^2,
\end{equation}
where \ $\overline f$ \ denotes the ensemble mean. Location
parameters \ $a_0\in{\mathbb R}, \ a_1, \ldots, a_M\geq 0$ \ and
scale parameters \ $b_0,b_1\geq 0$ \ can be estimated from the
training data by minimizing an appropriate verification score (see
Section \ref{subs:subs3.3}). 

In the case when the ensemble can be divided into groups of
exchangeable members, ensemble members within a given group get the
same coefficient of the  
location parameter resulting in a predictive distribution of the form
\begin{equation}
   \label{eq:eq3.7}
  {\mathcal N}\bigg(a_0+a_1\sum_{\ell_1=1}^{M_1}f_{\ell_1,1}+ \ldots
  +a_m\sum_{\ell_m=1}^{M_m} f_{\ell_m,m},b_0+b_1S^2\bigg), 
\end{equation}
where again, \ $S^2$ \ denotes the ensemble variance.

\subsubsection*{Wind speed}
To take into account the non-negativity of the predictable quantity, in
the EMOS  model for wind speed proposed by
\citet{tg}, the normal predictive distribution of
\eqref{eq:eq3.6} and \eqref{eq:eq3.7} is replaced by a truncated normal
distribution with cut-off at zero. In this way for exchangeable
ensemble members the predictive distribution is
\begin{equation}
   \label{eq:eq3.8}
  {\mathcal N}^{\,0}\bigg(a_0+a_1\sum_{\ell_1=1}^{M_1}f_{\ell_1,1}+ \ldots
  +a_m\sum_{\ell_m=1}^{M_m} f_{\ell_m,m},b_0+b_1S^2\bigg).
\end{equation}

A summary of the above described models is given in Table
\ref{tab:tab1}, where the 
BMA component and EMOS predictive PDFs and their mean/location and standard
deviation/scale parameters are given as functions of the ensemble members
$f_{\ell}$ and ensemble variance $S^2$. 

\begin{table}[t!]
\begin{center}
\begin{tabular}{|l|l|l|c|c|} \hline
& &Predictive PDF&Mean/location&Std. dev./scale\\ \hline
Temperature&BMA&Normal mixture&$\beta_{0,k}+\beta_{1,k}f_k$&$\sigma$\\
&EMOS&Normal&$a_{0}+\sum_{\ell=1}^Ma_{\ell}f_{\ell}$&$\sqrt{b_0+b_1S^2}$\\
\hline
&BMA&Gamma mixture&$b_0+b_1f_k$&$c_0+c_1f_k$\\
Wind speed&BMA&Truncated normal mixture&$\beta_{0,k}+\beta_{1,k}f_k$&$\sigma$\\
&EMOS&Truncated
normal&$a_{0}+\sum_{\ell=1}^Ma_{\ell}f_{\ell}$&$\sqrt{b_0+b_1S^2}$\\
\hline 
\end{tabular}
\caption{Summary of post-processing methods for temperature and wind
  speed forecasts. BMA component and
  EMOS predictive PDFs and their mean/location and standard
  deviation/scale parameters as functions of the ensemble members
  $f_{\ell}$ and ensemble variance $S^2$.} \label{tab:tab1} 
\end{center}
\end{table}

\subsection{Verification scores}
  \label{subs:subs3.3}

In order to check the overall performance of the calibrated 
forecasts in terms of probability
distribution function the mean continuous ranked probability scores 
\citep[CRPS;][]{wilks,grjasa} and the coverage and average width of 
 $83.33\,\%$ 
central prediction intervals are computed and compared for the
calibrated and raw ensemble.
Additionally, the ensemble mean and median
are used to consider point forecasts, which are evaluated with the use
of mean absolute errors (MAE) and root mean square errors (RMSE). 
We remark that for MAE and RMSE
the optimal point forecasts are the median and the mean, respectively
\citep{gneiting11, pinhag}.  Further, given a cumulative distribution
function (CDF) \ $F(y)$ \ and a real number \ $x$, \ the CRPS is defined as
\begin{equation*}
\crps\big(F,x\big):=\int_{-\infty}^{\infty}\big (F(y)-{\mathbbm 
  1}_{\{y \geq x\}}\big )^2{\mathrm d}y.
\end{equation*}
The mean CRPS of a probability forecast is the average of the CRPS values
of the predictive CDFs and corresponding validating observations taken
over all locations and time points considered. For the raw ensemble
the empirical CDF of the ensemble replaces the predictive CDF. Note
that CRPS, MAE and RMSE are negatively oriented scores, that is
the smaller the better. Finally, the
coverage of a \ $(1-\alpha)100 \,\%, \ \alpha \in (0,1),$ \ central prediction
interval is the proportion 
of validating observations located between the lower and upper \
$\alpha/2$ \ quantiles of the predictive distribution. For a
calibrated predictive PDF this value should be around \ $(1-\alpha)100
\,\%$.

\section{Results}
  \label{sec:sec4}
Using the ideas of \citet{bhn1,bhn2} we consider two different
groupings of the members of the ALADIN-HUNEPS ensemble. In the first
case we have two exchangeable groups \ ($m=2$). \  One contains the
control member 
denoted by \ $f_c \ (M_1=1)$, \ while in the other are 10 ensemble members
\ ($M_2=10$) \  corresponding to the differently perturbed initial
conditions denoted by 
\ $f_{p,1}, \ldots ,f_{p,10}$. \ Under these conditions for temperature data we
investigate BMA model \eqref{eq:eq3.3} with three different bias
correction methods (linear, additive, no bias correction) and EMOS
model \eqref{eq:eq3.7}, while for wind speed data BMA models
\eqref{eq:eq3.4} and \eqref{eq:eq3.5} and EMOS model  \eqref{eq:eq3.8}
are studied. In this two-group situation we have only one independent BMA
weight \ $\omega\in [0,1]$ \ corresponding e.g. to the control, that is \
$\omega_1=\omega$ \ and \ $\omega_2=(1-\omega)/10$. \

In the second case the odd and even numbered exchangeable ensemble
members form two separate groups $\{f_{p,1}, \ 
f_{p,3}, \ f_{p,5}, \ f_{p,7}, \ f_{p,9}\}$ \ and \ $\{f_{p,2}, \
f_{p,4}, \ f_{p,6}, \ f_{p,8}, \ f_{p,10}\}$, \ respectively \ ($m=3,
\ M_1=1, \ M_2=M_3=5$), \  which idea is
justified by the method their initial conditions
are generated. For more details see Section \ref{sec:sec2},
particularly the fact that only five perturbations are calculated and 
then they are added to (odd numbered members) and
subtracted from (even numbered members) the unperturbed
initial conditions. For calibrating ensemble
forecasts of temperature and wind speed we use the three-group
versions of BMA and EMOS models considered earlier in the two-group case.

\begin{figure}[t!]
\begin{center}
\leavevmode
\hbox{
\epsfig{file=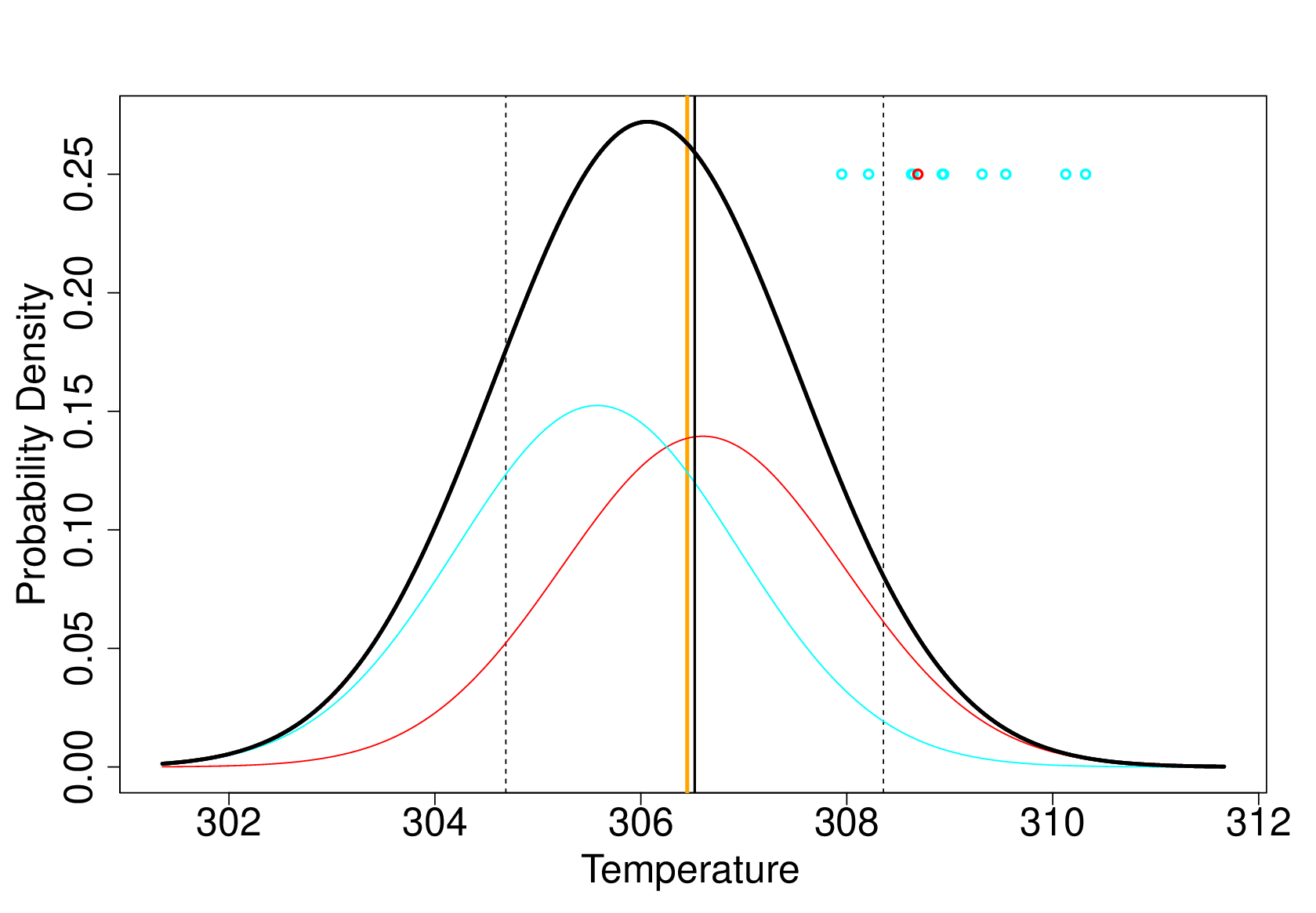,height=8cm, angle=-90} \quad
\epsfig{file=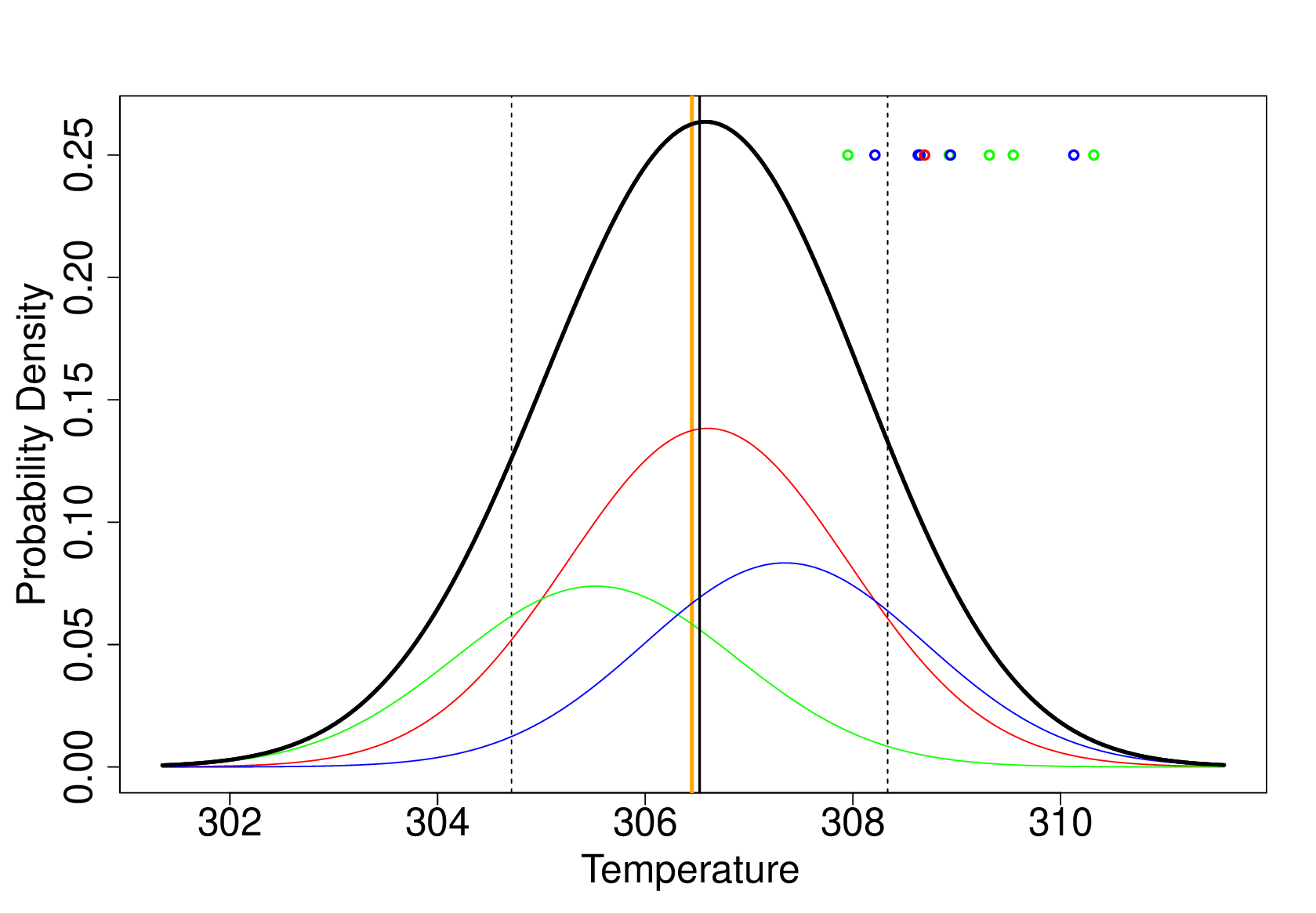,height=8cm, angle=-90}}

\centerline{\hbox to 9 truecm {\scriptsize (a) \hfill (b)}}

\hbox{
\epsfig{file=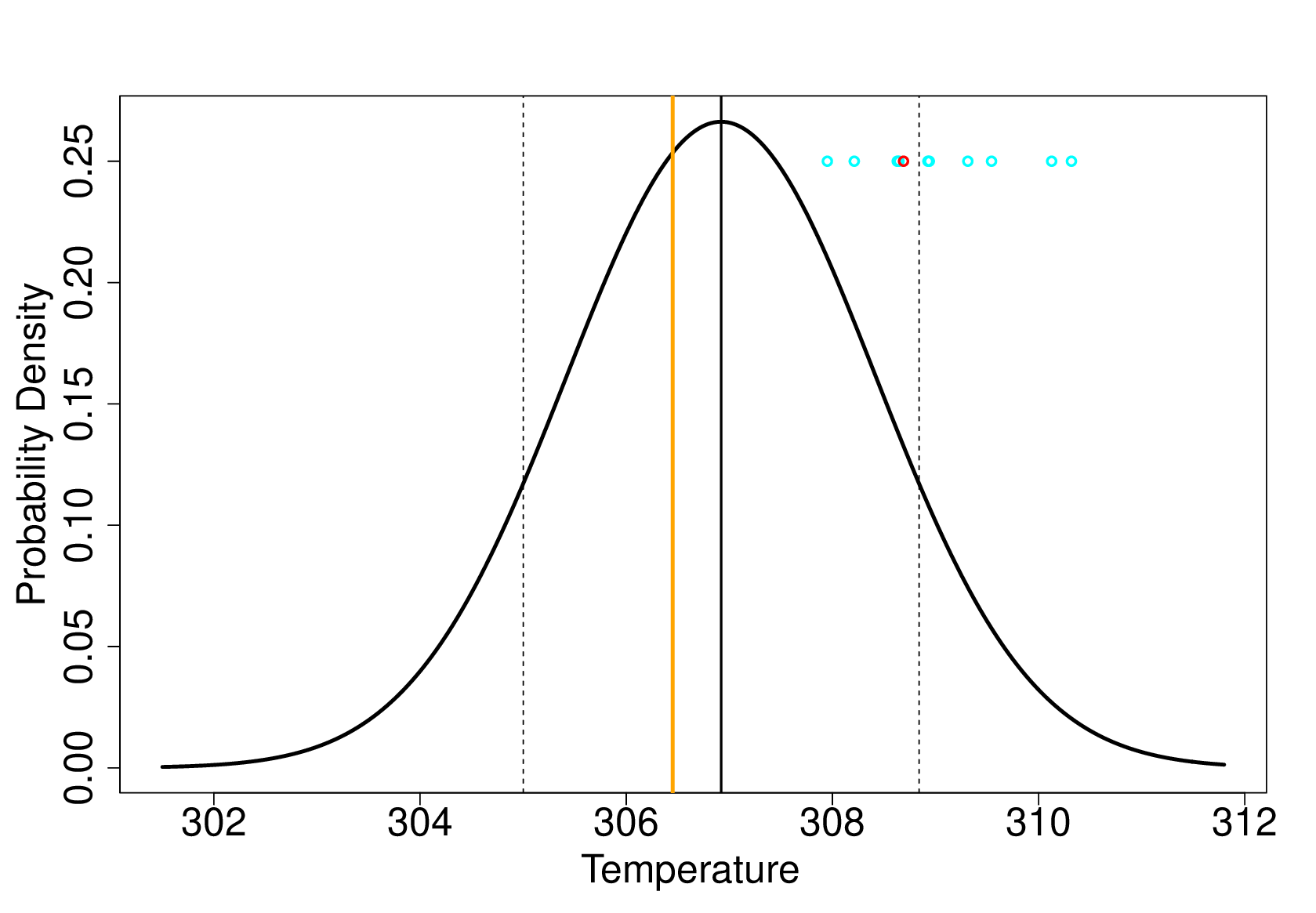,height=8cm, angle=-90} \quad
\epsfig{file=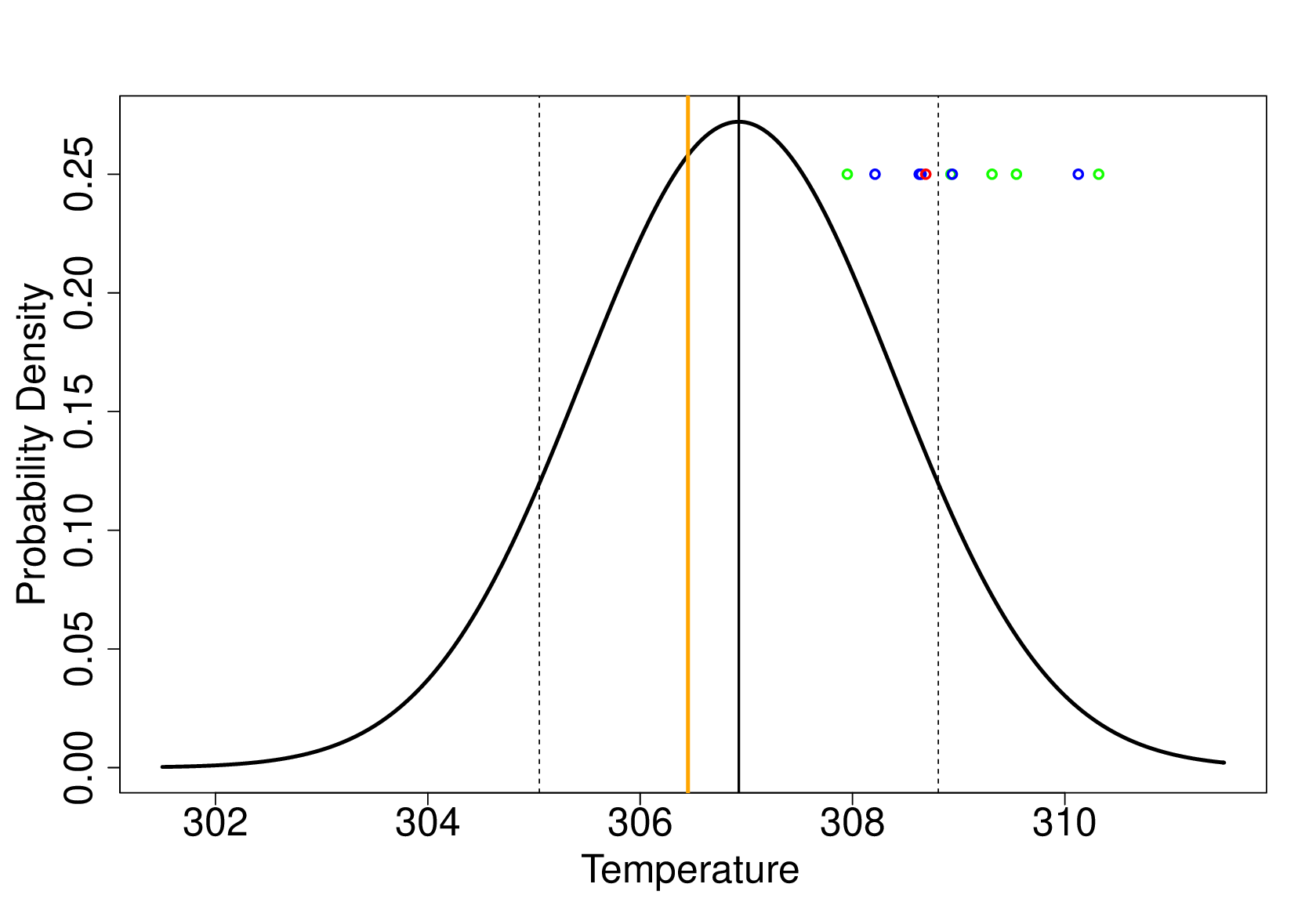,height=8cm, angle=-90}}

\centerline{\hbox to 9 truecm {\scriptsize (c) \hfill (d)}}
\caption{(a) and (b): BMA; (c) and (d): EMOS density forecasts for 2m
  temperature for Debrecen valid on 02.07.2012.
  BMA PDFs with linear bias correction in two- and three-group cases
  (overall: thick  black line; control: red 
  line; sum of exchangeable members on (a): light blue line;
  on (b): green (odd members) and blue (even members) lines), EMOS
  predictive PDFs in two- and three-group cases (thick black line),
  ensemble members (circles with the same colours as the corresponding
  BMA PDFs), BMA/EMOS median forecasts (vertical black line),
  verifying observations (vertical orange line) and the first and last
deciles (vertical dashed lines).}  
\label{fig:fig2}
\end{center}
\end{figure}

As typical example for illustrating the two different post-processing
methods and 
groupings we consider temperature data and forecasts for 
Debrecen valid on 02.07.2012.
Figures \ref{fig:fig2}a and \ref{fig:fig2}b show the BMA predictive
PDFs in the two- and in the three-group case, the component PDFs
corresponding to different groups, the median forecasts,
the verifying observations, the first and last deciles and the
ensemble members. Besides the EMOS predictive PDFs the same quantities
can be seen in Figures \ref{fig:fig2}c and \ref{fig:fig2}d, too.
On the considered date the spread of the ensemble
is reasonable (the ensemble range equals 2.368 K) but all
ensemble members overestimate the validating observation (306.45
K). Obviously the same holds for the ensemble 
median (308.927 K), while BMA median forecasts corresponding to the two- and
three-group models (both equal to 306.524 K) are
quite close to the true temperature. The point forecasts produced by
the EMOS model are slightly worse (306.921
K for both groupings) but still outperform the ensemble median.

\begin{figure}[t!]
\begin{center}
\leavevmode
\epsfig{file=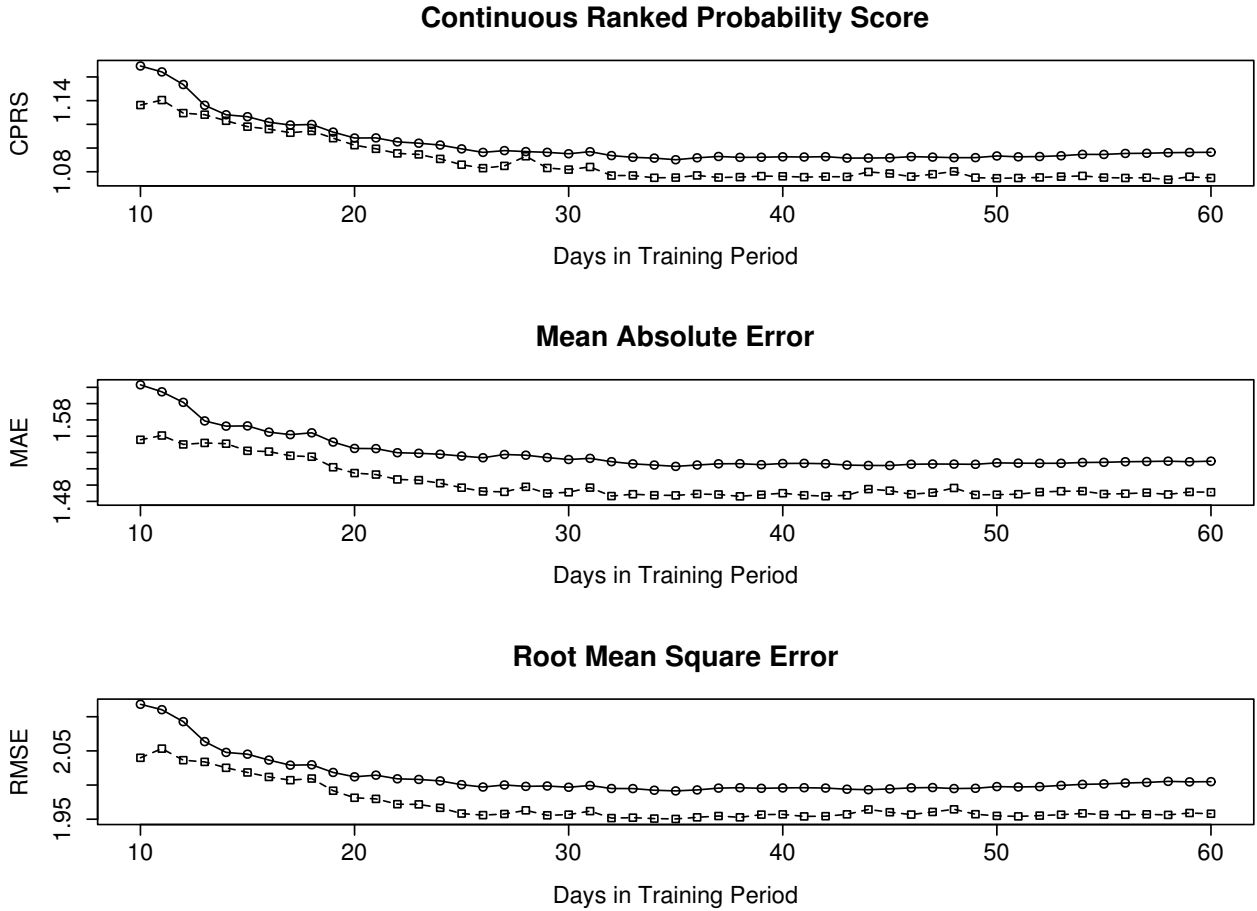,height=12cm, angle=0}
\caption{Mean CRPS of predictive distributions, MAE of
  BMA/EMOS median and RMSE of BMA/EMOS mean forecasts for temperature
  corresponding to two-group models for various training period
  lengths (BMA: solid and $\circ$; EMOS: dashed and $\Box$).}    
\label{fig:fig3}
\end{center}
\end{figure}

We start our data analysis by determining the optimal lengths of the
training periods to be used for estimating the parameters of BMA and EMOS
predictive distributions for 2m temperature and 10m wind speed. After
finding them we compare the performances of BMA and 
EMOS post-processed forecasts
using these optimal training period lengths. We remark that for EMOS models the
parameter estimates are obtained by minimizing the CRPS values of the
predictive PDFs. 

\subsection{Training period}
  \label{subs:subs4.1}

Similarly to our previous studies \citep{bhn1,bhn2} we proceed  in
the same way as \citet{rgbp} and determine the length of training
period to be used for BMA and EMOS calibrations by comparing the MAE
values of median
forecasts, the RMSE values of  mean forecasts, the CRPS values
of predictive distributions and the coverages and average
widths of  $83.33\,\%$ central prediction intervals calculated from
the predictive PDFs using training periods of length  $10,11,
\ldots ,60$  calendar days. In order to ensure the comparability of
the verification scores corresponding to different training period
lengths we issue calibrated forecasts of temperature and wind speed
for the period 01.06.2012 -- 31.03.2013 (6 days with missing data are
excluded). This means 298 calendar days following the first training
period of maximal length of 60 days.

\subsubsection*{Temperature}
For temperature data we consider BMA predictive PDF \eqref{eq:eq3.3} with
linear bias correction and EMOS  model \eqref{eq:eq3.7} with
parameters minimizing the CRPS of 
probabilistic forecasts corresponding to the training data. We remark 
that in order to ensure a more direct 
comparison of the two models we also investigated the performance of
the BMA predictive PDF \eqref{eq:eq3.3} 
with parameter estimates minimizing the same verification
score. It yielded sharper central prediction intervals and lower
coverage for all training period length considered, but there were no 
significant differences in CRPS, MAE and RMSE values corresponding to
different parameter estimation methods. 

\begin{table}[t!]
\begin{center}
\begin{tabular}{|l|l|c|c|c|c|c|c|c|c|c|} \hline
\multicolumn{2}{|c|}{}&\multicolumn{3}{|c|}{Mean CRPS}&
\multicolumn{3}{|c|}{MAE}&
\multicolumn{3}{|c|}{RMSE}\\ \cline{3-11}
\multicolumn{2}{|c|}{}&opt.&opt.&day 35 &opt.&opt.&day
35&opt.&opt.&day 35
\\
\multicolumn{2}{|c|}{}&day&value&value&day&value&value&day&value&value
\\ \hline 
Two
&BMA&$35$&$1.0900$&$1.0900$&$35$&$1.5230$&$1.5230$&$35$&$1.9914$&$1.9914$
\\ 
groups&EMOS&$58$&$1.0734$&$1.0751$&$38$&$1.4862$&$1.4874$&$35$&$1.9504$&$
1.9504$ \\ \hline
Three
&BMA&$35$&$1.0896$&$1.0896$&$35$&$1.5227$&$1.5227$&$36$&$1.9897$&$1.9899$
\\
groups&EMOS&$36$&$1.0747$&$1.0756$&$28$&$1.4821$&$1.4901$&$28$&$1.9506$
&$1.9576$\\
\hline  
\end{tabular} 
\caption{Optimal training period lengths for temperature with respect
  to mean CRPS, MAE and RMSE, the corresponding optimal scores and
  scores at the chosen 35 days length. 
} \label{tab:tab2}
\end{center}
\end{table}
Consider first the two-group situation. In Figure \ref{fig:fig3} the
CRPS values of BMA and EMOS predictive distributions, MAE values of
median and RMSE values of mean forecasts are plotted against the
length of the training period. We remark that for normal EMOS
model mean and median forecasts are obviously coincide. 
First of all it is noticeable that the results are very consistent for
all diagnostics, i.e. the curves are similar for all measures. EMOS
produces better verification scores and after 32 days there is no
big difference among scores obtained with different training period
lengths. In case of the BMA model CRPS, MAE and RMSE reach their minima 
at day 35 and this training period length gives the minimum of RMSE  
of the EMOS model, too (see Table \ref{tab:tab2}). Although the minima
of CRPS and MAE  
of the EMOS model are reached at days 58 and 38, respectively, 
the values at day 35 are very near to these minima as well.

\begin{figure}[t!]
\begin{center}
\leavevmode
\epsfig{file=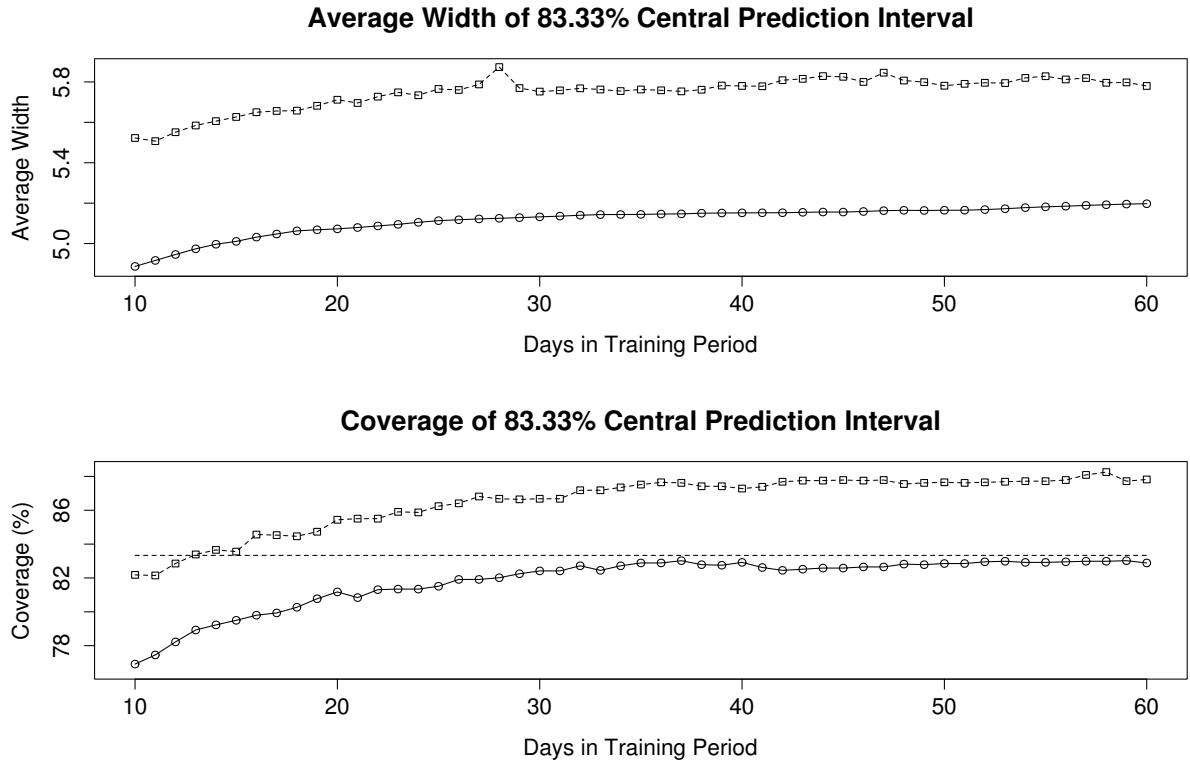,height=10cm, angle=0}
\caption{Average width and coverage of $83.33\,\%$ BMA/EMOS central prediction
  intervals for temperature corresponding to two-group models for
  various training period lengths (BMA: solid and $\circ$; EMOS:
  dashed and $\Box$).}      
\label{fig:fig4}
\end{center}
\end{figure}
Figure \ref{fig:fig4} shows the average width and the
coverage of the  $83.33\,\%$ central prediction interval for both
models considered. Similarly to the
previous diagnostics, after 32 days all curves are rather
flat showing only a slightly increasing trend. EMOS model yields
significantly wider central prediction 
intervals for all training period lengths considered and
its coverage is above the nominal value of $83.33\,\%$ (dashed line) for 
training periods longer than 12 days. Unfortunately, the coverage of
the BMA model fails to reach the nominal value, but it is very close to
$83.33\,\%$  from day 35 onwards. The maximal coverage 
is attained at day 37. Comparing the average width and coverage one can observe
that they have opposite behaviour, i.e. the average width values
favour shorter training periods, while 
the coverage figures prefer longer ones. On the other hand, the trend
of the average width values is rather flat after day 30 (or so). For
any case a reasonable compromise ought to be found, which is at the
range of 30 - 40 days.   

As a summary it can be said that a 35 days training period seems to be
an acceptable choice 
both for the BMA  and for the EMOS models (particularly 
see conclusions based on Figure \ref{fig:fig3}, which are not compromised by
the other two diagnostics at Figure \ref{fig:fig4}). 

\begin{figure}[t!]
\begin{center}
\leavevmode
\epsfig{file=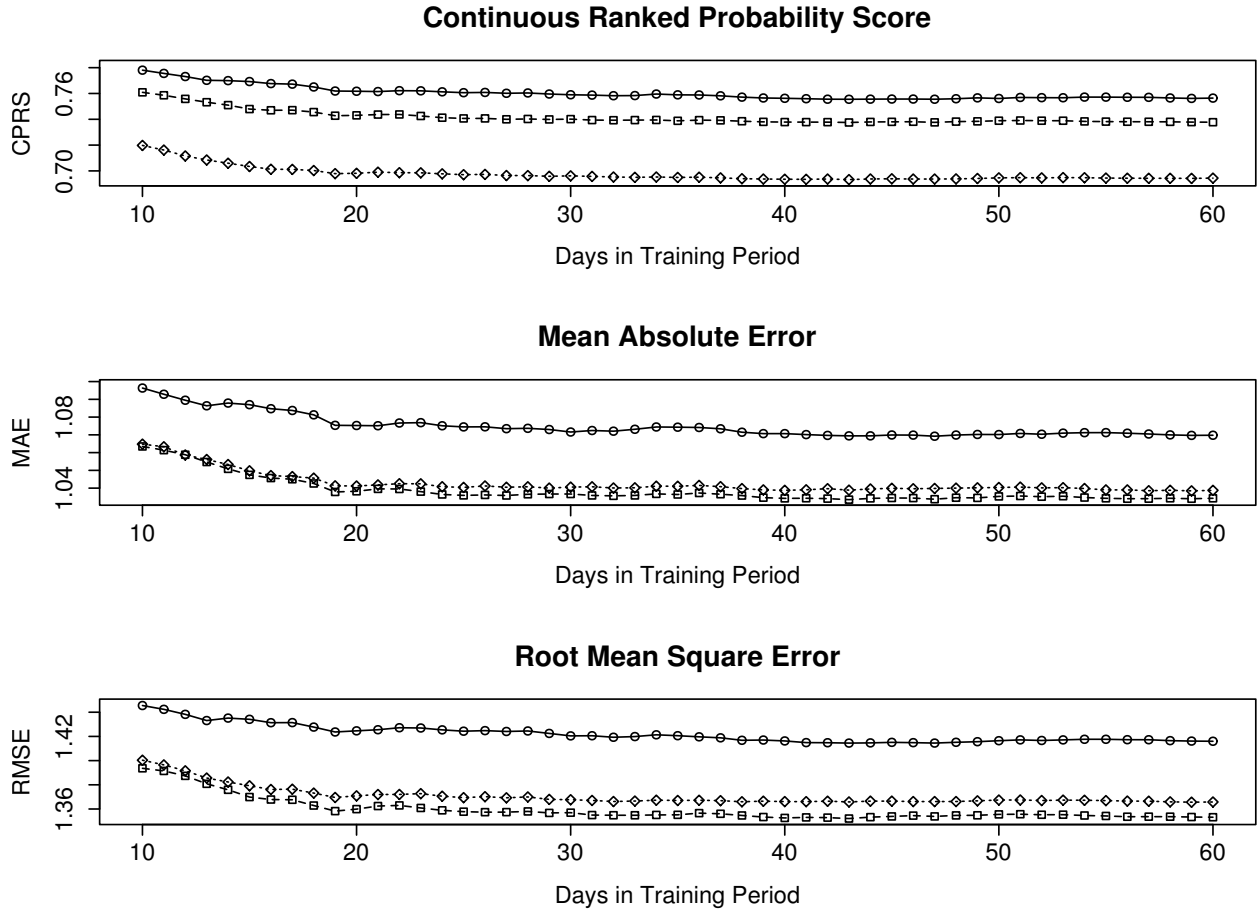,height=12cm, angle=0}
\caption{Mean CRPS of predictive distributions, MAE of
  BMA/EMOS median and RMSE of BMA/EMOS mean forecasts for wind speed
  corresponding to two-group models for various training period
  lengths (Gamma BMA: solid and $\circ$; truncated normal BMA: dotted
  and $\diamond$; EMOS: dashed and $\Box$).}    
\label{fig:fig5}
\end{center}
\end{figure}
Very similar conclusions can be drawn 
for the three-group models. The overall behaviours of the two
post-processing methods for the various diagnostics (not shown) is
very similar to those of their two-group counterparts. EMOS model provides
lower CRPS, MAE 
and RMSE values and higher coverage combined with wider central
prediction interval all over the time periods. In terms of specific
values the minima of CRPS and MAE 
for the BMA model are reached again at day 35, while the RMSE takes its minimum
at day 36 (the value at day 35 is very near to this
minimum, see Table \ref{tab:tab2}). For the EMOS model CRPS, MAE and
RMSE reach their 
minima in the range of 28 - 36 days and values at day 35 are
again very close to these minima.

\begin{figure}[t!]
\begin{center}
\leavevmode
\epsfig{file=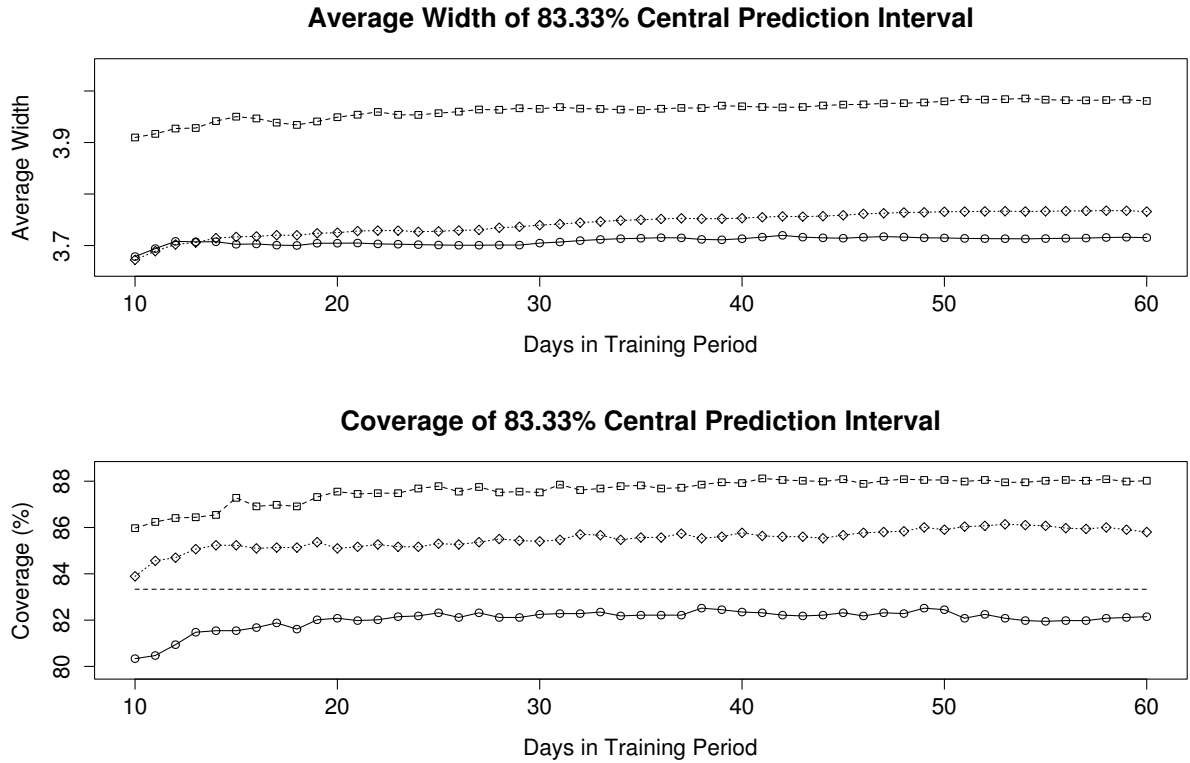,height=10cm, angle=0}
\caption{Average width and coverage of $83.33\,\%$ BMA/EMOS central prediction
  intervals for wind speed corresponding to two-group models for
  various training period lengths (Gamma BMA: solid and $\circ$;
  truncated normal BMA: dotted 
  and $\diamond$; EMOS: dashed and $\Box$).}     
\label{fig:fig6}
\end{center}
\end{figure}

Regarding the average width, shorter training periods yields sharper
central prediction intervals. The coverage for the EMOS model is above
the nominal value for training periods longer than 14 days, while the
maximal coverage of the BMA model is reached at day 59. However, as in general
shorter training periods are preferred, a reasonable compromise is to
consider the 35 - 38 days interval where the BMA coverage is
also very high.  Hence, the training period 
proposed for the two-group model can be kept for the three-group
model as well, therefore for temperature we suggest the  
use of a training period of length 35 days for all the investigated
post-processing methods.

\subsubsection*{Wind speed}
To calibrate ensemble forecasts of wind speed we apply gamma BMA model
\eqref{eq:eq3.4}, truncated normal BMA model \eqref{eq:eq3.5} and EMOS
model \eqref{eq:eq3.8}. In the latter case, similarly to EMOS
calibration of temperature forecasts, estimation of parameters is done
by minimizing the CRPS of probabilistic forecasts corresponding to the
training data.

First, consider again the case when we have two groups of exchangeable
ensemble members. Generally the various scores have rather flat
evolution with respect to the training lengths (see Figures
\ref{fig:fig5} and \ref{fig:fig6}). It is particularly true after day
25, which would suggest that basically any training length longer than
25 days might be an acceptable choice. Observe that the order of
different methods with respect to a given score remains the same for
all training period lengths. Truncated normal BMA produces the lowest CRPS
values, while the best MAE and RMSE values correspond to EMOS
post-processing. For any case if we wanted to pick up a single training
period length as an optimal one, 43 days would be a reasonable choice. This is  
the value where the minima of CRPS of all three methods, the minima of
RMSE of gamma BMA and of EMOS and the minimum of
MAE of EMOS are reached (see Table \ref{tab:tab3}). The other two
scores corresponding 
to EMOS model attain their minima at day 59, however values corresponding
to day 43 are practically the same. Finally, the minimum of  MAE
of the gamma BMA model is reached at day 47, while
the value at day 43 is the second smallest one.   

EMOS post-processing yields the widest central
prediction intervals and the highest coverage (Figure \ref{fig:fig6})
which is above the nominal level for all considered training period
lengths. The $83.33\,\%$ central prediction intervals corresponding to
the truncated
normal BMA model are significantly sharper than those of the EMOS combined
with a coverage varying between $83.89\,\%$ and $86.14\,\%$. Gamma BMA
results in even narrower central prediction intervals but its coverage
never reaches the nominal level. The maximal coverage 
is attained at days 38 and 49. In this way a 43 days training period
length is also acceptable from the point of view of central prediction
intervals. 

\begin{table}[t!]
\begin{center}
\begin{tabular}{|l|l|c|c|c|c|c|c|c|c|c|} \hline
\multicolumn{2}{|c|}{}&\multicolumn{3}{|c|}{Mean CRPS}&
\multicolumn{3}{|c|}{MAE}&
\multicolumn{3}{|c|}{RMSE}\\ \cline{3-11}
\multicolumn{2}{|c|}{}&opt.&opt.&day 43 &opt.&opt.&day
43&opt.&opt.&day 43
\\
\multicolumn{2}{|c|}{}&day&value&value&day&value&value&day&value&value
\\ \hline 
Two
&BMA, g.&$43$&$0.7551$&$0.7551$&$47$&$1.0692$&$1.0694$&$43$&$1.4145$&$1.4145$
\\ 
groups
&BMA, tr.&$43$&$0.6933$&$0.6933$&$59$&$1.0385$&$1.0389$&$59$&$1.3657$&$1.3658$
\\ 
&EMOS&$43$&$0.7375$&$0.7375$&$43$&$1.0335$&$1.0335$&$43$&$1.3521$&$
1.3521$ \\ \hline
Three
&BMA, g.&$43$&$0.7559$&$0.7559$&$42$&$1.0690$&$1.0691$&$42$&$1.3940$&$1.3941$
\\
groups
&BMA, tr.&$39$&$0.6930$&$0.6930$&$39$&$1.0377$&$1.0382$&$39$&$1.3545$&$1.3551$
\\
&EMOS&$43$&$0.7393$&$0.7393$&$43$&$1.0342$&$1.0342$&$43$&$1.3540$
&$1.3540$\\
\hline  
\end{tabular} 
\caption{Optimal training period lengths for wind speed with respect
  to mean CRPS, MAE and RMSE, the corresponding optimal scores and
  scores at the chosen 43 days length. 
} \label{tab:tab3}
\end{center}
\end{table}
The analysis of verification scores corresponding to the alternative
grouping of ensemble members (not shown) leads again to very similar
results. The most important 
difference between the two-group and three-group models is that
forming three groups (especially for training periods
longer than 20 days) improves MAE and RMSE values of
the truncated normal BMA model and they become very close to the
corresponding values of EMOS. For the three-group EMOS model all
three verification scores reach their minima at day 43 and this is the training
period length where the minimal CRPS  and the second
smallest values of MAE and RMSE of the gamma BMA model are
attained (see Table \ref{tab:tab3}). For the latter model the global
minima of MAE 
and RMSE are at day 42. In case of truncated normal BMA
post-processing CRPS, MAE and RMSE have their minima at day 39, but
since these curves are rather flat, values corresponding to a training period of
length 43 days are very near.
In this way a 43 days training period seems to be
acceptable for both groupings of ensemble members.

\subsection{Ensemble calibration using BMA and EMOS post-processing}     
  \label{subs:subs4.2}

According to the results of the previous section to compare the
performance of BMA and EMOS post-processing on the 11 member
ALADIN-HUNEPS ensemble we use rolling training periods of lengths 35
days for temperature and 43 days for wind speed. 

\subsubsection*{Temperature}
For post-processing ensemble forecasts of temperature we consider the
BMA model \eqref{eq:eq3.3} with all three bias correction methods
introduced in Section \ref{subs:subs3.1} (linear, additive, none)
and EMOS model minimizing the CRPS of probabilistic forecasts
corresponding to the training data. The application of three different BMA bias
correction methods is justified by a previous study dealing with
statistical calibration of the ALADIN-HUNEPS temperature forecasts
\citep{bhn2}, where the simplest BMA model without bias correction showed the
best overall performance (although that study was using different
ALADIN-HUNEPS dataset period, which preceded the one investigated in
this article). 

The use of a 35 day rolling training period implies that ensemble members,
validating observations and predictive PDFs are available for the
period  07.05.2012 -- 31.03.2013 (having 323 calendar days just after
the first 35 days training period). This time interval starts nearly 4
weeks earlier than the one used for determination of the optimal
training period length. 

\begin{figure}[t!]
\begin{center}
\leavevmode
\epsfig{file=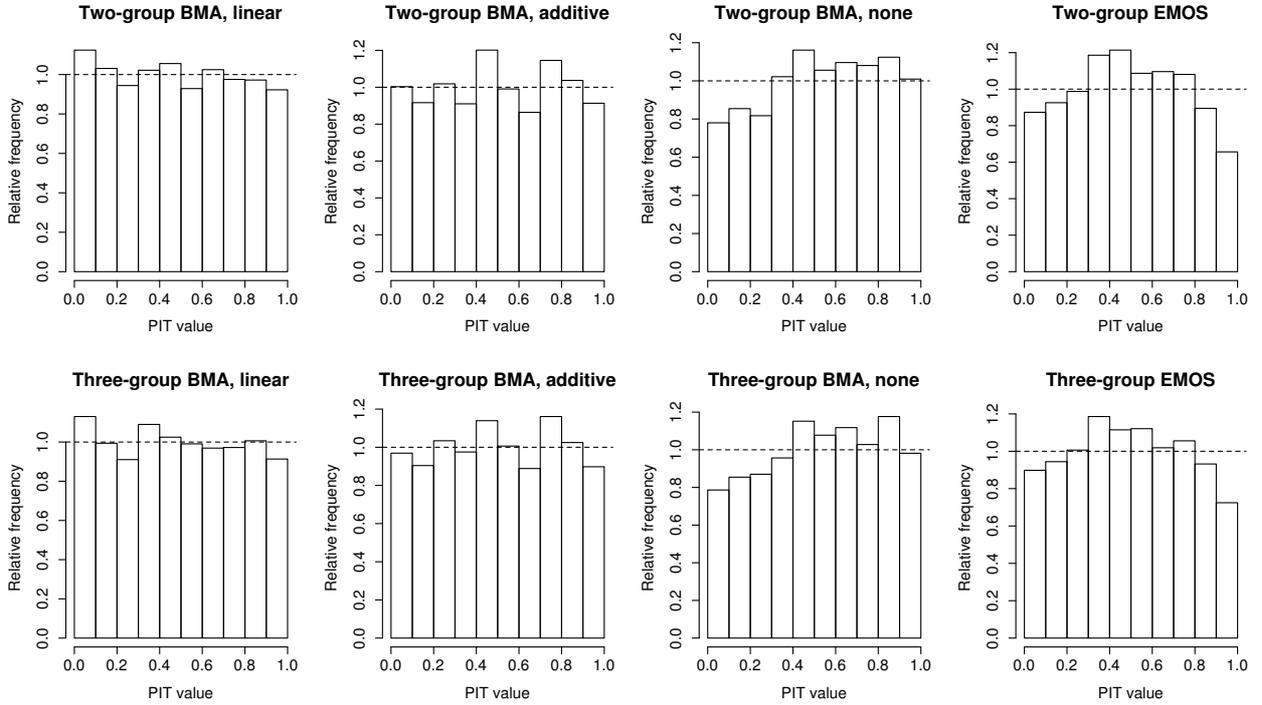,height=9.2cm}
\caption{PIT histograms for BMA and EMOS post-processed forecasts of
  temperature using two- and three-group models.}   
\label{fig:fig7}
\end{center}
\end{figure} 

The first step in checking the calibration of our post-processed
forecasts is to have a look at their probability integral transform
(PIT) histograms. The PIT is the value of the predictive cumulative
distribution evaluated at the verifying observations \citep{rgbp},
which provides a good and easily interpretable measure about the
possible improvements of the under-dispersive character of the raw
ensemble. The closer the 
histogram to the uniform distribution, the better the calibration
is. In Figure \ref{fig:fig7} PIT histograms corresponding to all three
versions of the BMA model and to the EMOS model are displayed both in the
two- and in the three-group case. A comparison to
the verification rank histogram of the raw ensemble (see Figure
\ref{fig:fig1}) shows that at every case post-processing significantly
improves the statistical calibration of the forecasts. However, the BMA 
model without bias correction and the EMOS model now become
slightly over-dispersive, while at the same time for the BMA models
with linear and additive bias correction one can
accept uniformity. This visual perception is confirmed by the $p$-values
of Kolmogorov-Smirnov tests for uniformity of the PIT values (see Table
\ref{tab:tab4}). Therefore, the BMA model with additive bias
correction produces the best PIT histograms (the linear bias
correction case is just slightly 
worse), while the fit of the BMA model without bias correction and
that of the EMOS model are rather poor. One can additionally observe
that the three-group models systematically outperform the two-group ones.

\begin{table}[b!]
\begin{center}
\begin{tabular}{|l|c|c|c|c|} \hline
&\multicolumn{3}{|c|}{BMA model with bias correction}&EMOS model \\ \cline{2-4}
&linear&additive&none& \\ \hline
Two groups&$0.1393$&$0.2405$&$2.2\times 10^{-10}$&$9.4\times 10^{-7}$\\
Three groups&$0.2281$&$0.4617$&$4.1\times 10^{-9}$&$8.6\times 10^{-5}$ \\ \hline 
\end{tabular} 
\caption{$p$-values of Kolmogorov-Smirnov tests for
  uniformity of PIT 
  values corresponding to predictive distributions of
  temperature.} \label{tab:tab4}       
\end{center}
\end{table}

\begin{table}[t!]
\begin{center}
\begin{tabular}{|l|l|c|c|c|c|c|c|c|c|c|c|c|c|} \hline
\multicolumn{2}{|c|}{}&Mean&
\multicolumn{2}{|c|}{MAE}&
\multicolumn{2}{|c|}{RMSE}&Average&Coverage\\ \cline{4-7}
\multicolumn{2}{|c|}{}&CRPS&median&mean&median&mean&width&($\%$)
\\ \hline 
&BMA, lin.&$1.0815$&$1.5101$&$1.5097$&$1.9789$&$1.9765$&$5.1375$&$83.00$
\\ 
Two
&BMA, add.&$1.1029$&$1.5395$&$1.5329$&$2.0028$&$1.9871$&$5.5146$&$84.21$
\\ 
groups&BMA, none&$1.1131$&$1.5536$&$1.5444$&$2.0167$&$2.0014$&$5.7191$&$84.80$
\\
&EMOS&$1.0693$&$1.4768$&$1.4768$&$1.9387$&$1.9387$&$5.7735$&$87.55$
\\ \hline
&BMA, lin.&$1.0801$&$1.5082$&$1.5059$&$1.9767$&$1.9726$&$5.1369$&$83.28$\\
Three
&BMA, add.&$1.0998$&$1.5346$&$1.5254$&$1.9962$&$1.9783$&$5.5096$&$84.12$
\\
groups&BMA, none&$1.1123$&$1.5509$&$1.5407$&$2.0156$&$1.9988$&$5.7095$&$85.11$
\\
&EMOS&$1.0670$&$1.4743$&$1.4743$&$1.9399$&$1.9399$&$5.6143$&$86.07$
\\\hline 
\multicolumn{2}{|l|}{Raw
  ensemble}&$1.2284$&$1.5674$&$1.5512$&$2.0434$&$2.0131$&$3.9822$&$60.53$\\
\hline  
\end{tabular} 
\caption{Mean CRPS of probabilistic, MAE and RMSE of
median/mean forecasts, average width and coverage of $83.33\,\%$
central prediction intervals for temperature.} \label{tab:tab5}
\end{center}
\end{table}

In Table \ref{tab:tab5} verification measures of probabilistic and point
forecasts calculated using BMA and EMOS models are given together with
the corresponding scores of the raw ensemble. By examining these
results, one can clearly observe again the obvious advantage of 
post-processing with respect to the raw ensemble. This is quantified
in decrease of CRPS, MAE and RMSE values and in a significant increase
in the coverage of the 
$83.33\,\%$ central prediction intervals. On the other hand, the post-processed
forecasts are less sharp (e.g. $83.33\,\%$ central prediction
intervals are around $30 \, \% - 40\, \%$ wider than the raw ensemble
range). This fact is coming from the small dispersion of the raw ensemble, as
also seen in the verification rank histogram of Figure
\ref{fig:fig1}. Furthermore, BMA and EMOS models distinguishing three
exchangeable groups of ensemble members slightly outperform their
two-group counterparts (in agreement with the interpretations based on the
PIT histograms). Comparing the different post-processing methods it is
noticeable that on the one hand EMOS produces the lowest CRPS, MAE and
RMSE values both in the two- and in the three-group case although with
coverages highly above the 
targeted $83.33\,\%$. On the other hand, in terms of CRPS, MAE and
RMSE the behaviour of 
the BMA model with linear bias correction is just slightly worse and at
the same time this method produces the sharpest predictive PDFs and the best
approximation of the nominal coverage. Taking also into account 
the fit of the PIT values to the uniform distribution (see Figure
\ref{fig:fig7} and Table \ref{tab:tab4} again) one can conclude
that overall from the four competing post-processing methods the BMA model with
linear bias correction shows the best performance. These results are not
in contradiction with the ones for a previous period (see \citet{bhn2}, where 
the no bias correction case proved to be the optimal) since the
characteristics of the raw ALADIN-HUNEPS system had been slightly
changed in between. The coverage of the system had been significantly
improved (from $46\,\%$ to $60\, \%$) although the latest system became slightly
biased (as compared to the previously examined one). Therefore, due to
the existence of the bias it is not surprising that one of the
versions with bias correction has the best behaviour.

\subsubsection*{Wind speed}

\begin{figure}[t!]
\begin{center}
\leavevmode
\epsfig{file=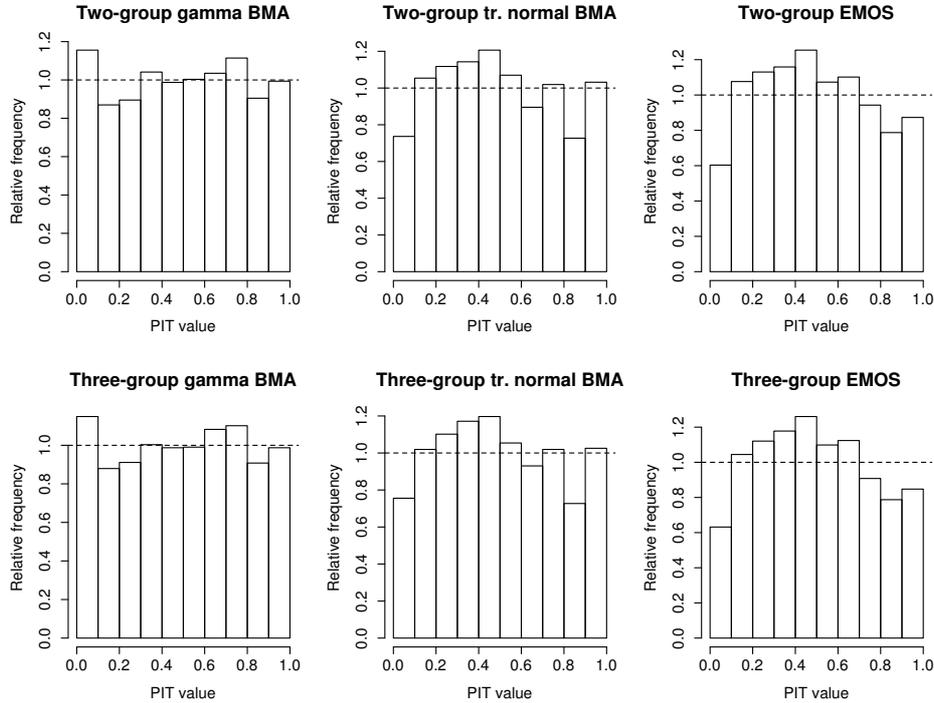,height=9.2cm}
\caption{PIT histograms for BMA and EMOS post-processed forecasts of
  wind speed using two- and three-group models.}   
\label{fig:fig8}
\end{center}
\end{figure}  
According to results of Section \ref{subs:subs4.1}, to compare the
performance of gamma BMA \eqref{eq:eq3.4}, truncated normal BMA
\eqref{eq:eq3.5} and EMOS \eqref{eq:eq3.8} post-processing on the 11
member ALADIN-HUNEPS ensemble we use a training period 
of length 43 calendar days. In this way ensemble members, validating
observations and predictive distributions are available for the period
15.05.2012 -- 31.03.2013 (313 calendar days).

\begin{table}[t!]
\begin{center}
\begin{tabular}{|l|c|c|c|} \hline
&\multicolumn{2}{|c|}{BMA model}&EMOS model \\ \cline{2-3}
&gamma&tr. normal& \\ \hline
Two groups&$0.1812$&$0.0023$&$1.8\times 10^{-5}$\\
Three groups&$0.2085$&$0.0043$&$9.6\times 10^{-7}$ \\ \hline 
\end{tabular} 
\caption{$p$-values of Kolmogorov-Smirnov tests for
  uniformity of PIT 
  values corresponding to predictive distributions of
  wind speed.} \label{tab:tab6}       
\end{center}
\end{table}
First, consider again the PIT histograms of various
calibration methods, which are displayed in Figure \ref{fig:fig8}. Compared to
the verification rank histogram of the wind speed ensemble (see Figure
\ref{fig:fig1}) the statistical post-processing induced improvements
are obvious, however, e.g. in case of EMOS both corresponding PIT
histograms are highly over-dispersive. In that sense the temperature
calibration (as shown in the previous chapter) is more successful than
the wind speed one. The $p$-values of 
Kolmogorov-Smirnov tests given in Table \ref{tab:tab6} also show that EMOS
models produce the poorest fit, while for gamma BMA models one can
accept uniformity. In case of BMA calibration the three-group models
again outperform the
two-group ones, while for EMOS the situation is the reverse. A similar
behaviour can be observed  in Table \ref{tab:tab7}, where the
verification scores of probabilistic and point 
forecasts calculated using BMA and EMOS post-processing and the corresponding
measures of the raw ensemble are given. 
Considering first the probabilistic forecasts (in terms of CRPS,
average width of central prediction interval and coverage) one can
observe that the calibrated forecasts are smaller in CRPS, wider in
central prediction intervals and
higher in coverage compared to the raw ensemble. Equally, to the two-
and in three-group cases the 
smallest CRPS values belong to the truncated normal BMA model, while
gamma BMA post-processing produces the best approximation of the
nominal coverage of $83.33\,\%$. Regarding the point forecasts (median and mean)
calculated from the truncated normal BMA and EMOS predictive PDFs, generally
there are smaller MAE and RMSE values than those of the raw
ensemble. However, there is an exception for the gamma BMA
model since these scores are higher indicating degradations. A possible
explanation might be related to the fact that in the investigated
period (15.05.2012 -- 31.03.2013) both  
the raw ensemble median and the ensemble mean slightly overestimate the
validating observations (their average biases (standard errors) are
$0.0907 \ (0.0249)$ and $0.0972 \ (0.0244)$,
respectively). Therefore, the small bias should be removed by
relevant bias corrections. On the other hand, we believe that the
simplest bias correction procedure of the gamma BMA model cannot eliminate these
inaccuracies, moreover, it might introduce some additional errors. It
is a matter of fact that in the
two-group case the average biases of the median and mean of the gamma BMA
predictive PDF are $-0.1935$ and $-0.1318$ with standard errors of
$0.0250$ and $0.0253$, respectively, while for the EMOS model showing the
lowest MAE and RMSE values these biases are only $-0.0340$ and $0.0312$, both
having a standard error of $0.0249$. Therefore, the EMOS model is able
to compensate for the existing biases, which is also the case for the
truncated normal BMA case, but not for the gamma BMA calibration. The
difference in behaviour between the two BMA calibration methods is
attributed to the more sophisticated bias correction algorithm, which
is applied for the truncated normal BMA case. 

\begin{table}[t!]
\begin{center}
\begin{tabular}{|l|l|c|c|c|c|c|c|c|c|c|c|c|c|} \hline
\multicolumn{2}{|c|}{}&Mean&
\multicolumn{2}{|c|}{MAE}&
\multicolumn{2}{|c|}{RMSE}&Average&Coverage\\ \cline{4-7}
\multicolumn{2}{|c|}{}&CRPS&median&mean&median&mean&width&($\%$)
\\ \hline 
Two
&BMA, g.&$0.7601$&$1.0747$&$1.0895$&$1.4176$&$1.4267$&$3.7151$&$81.87$
\\ 
groups&BMA, tr. n.&$0.6982$&$1.0446$&$1.0478$&$1.3693$&$1.3772$&$3.7621$&$85.46$
\\
&EMOS&$0.7409$&$1.0384$&$1.0451$&$1.3583$&$1.3602$&$3.9777$&$88.06$
\\ \hline
Three
&BMA, g.&$0.7612$&$1.0754$&$1.0828$&$1.4192$&$1.4052$&$3.7064$&$82.03$
\\
groups&BMA, tr. n.&$0.6980$&$1.0437$&$1.0449$&$1.3696$&$1.3649$&$3.7498$&$85.08$
\\
&EMOS&$0.7431$&$1.0396$&$1.0469$&$1.3597$&$1.3633$&$4.0294$&$88.25$
\\\hline 
\multicolumn{2}{|l|}{Raw
  ensemble}&$0.8029$&$1.0688$&$1.0549$&$1.3980$&$1.3728$&$2.8842$&$68.22$\\
\hline  
\end{tabular} 
\caption{Mean CRPS of probabilistic, MAE and RMSE of
median/mean forecasts, average width and coverage of $83.33\,\%$
central prediction intervals for wind speed.} \label{tab:tab7}
\end{center}
\end{table}
To summarize, gamma BMA model outperforms the other two methods in terms
of fit of PIT values and in sharpness 
and coverage of central prediction intervals, but it has the highest
CRPS and very poor verification scores for the point forecasts. MAE
and RMSE values 
corresponding to EMOS and truncated normal BMA are lower than those of
the raw ensemble and rather
similar to each other. From these two methods truncated normal BMA
produces much lower CRPS, sharper central 
prediction intervals and better fit of PIT values to the uniform
distribution, so we conclude that the overall performance
of this method is the best for the calibration of the wind speed raw
ensemble forecasts.

\section{Discussion and conclusions}
  \label{sec:sec5}
 
In this paper we have compared different versions of the BMA and EMOS
statistical post-processing methods in order to improve the
calibration of 2m temperature and 10m wind speed forecasts of the
ALADIN-HUNEPS system. First, we have demonstrated the weaknesses of the
ALADIN-HUNEPS raw ensemble system being under-dispersive and therefore
uncalibrated. We have indicated that the under-dispersive character of
the ALADIN-HUNEPS system had been improved compared to studies based
on a former dataset, however more enhancements are still needed. On
the other hand, the latest dataset shows some features of bias of
ALADIN-HUNEPS, which were inexistent in earlier studies. This fact has
an influence on the optimal choice of statistical calibration, since
the use of bias correction is getting more essential. Some standard measures
were applied, which are related to the characteristics of the ensemble
probability density functions and also the point forecasts as
described by the mean/median of the ensemble. The various systems
improve different aspects of the ensemble, however overall both the
BMA and the EMOS method is capable to deliver significant improvements
on the raw ALADIN-HUNEPS ensemble forecasts (for temperature and wind speed
as well). Generally the best BMA method slightly outperforms the EMOS
technique (although it should not be forgotten that for instance in
terms of point forecasts EMOS is better than BMA).  

\bigskip
\noindent
{\bf Acknowledgments.} \  \ Research was supported by 
the Hungarian  Scientific Research Fund under Grant No. OTKA NK101680
and by the T\'AMOP-4.2.2.C-11/1/KONV-2012-0001 
project. The project has been supported by the European Union,
co-financed by the European Social Fund. Essential part of this work
was made during the visiting professorship of the first author at the
Institute of Applied Mathematics of the University of Heidelberg.
The authors are indebted to Tilmann
Gneiting for his useful suggestions and
remarks, to Thordis Thorarinsdottir for the R codes of EMOS for wind
speed and to  Mih\'aly Sz\H ucs from the HMS for providing the 
data.


\begin{thebibliography}{99}
\bibitem[Baran, 2013]{bar} Baran, S. (2013) Probabilistic wind speed
  forecasting using Bayesian 
   model averaging with truncated normal components. {\em
     Comput. Stat. Data. Anal.\/},  under review (arXiv:1305:1184).

\bibitem[Baran {\em et al.\/}, 2013a]{bhn1}  Baran, S., Hor\'anyi, A. and
  Nemoda, D. (2013a) Statistical 
  post-processing of probabilistic wind speed
  forecasting in Hungary. {\em Meteorol. Z.\/} {\bf 22}, 273--282.

\bibitem[Baran {\em et al.\/}, 2013b]{bhn2}  Baran, S., Hor\'anyi, A. and
  Nemoda, D. (2013b) Probabilistic 
   temperature forecasting with statistical calibration in Hungary.  
  {\em Meteorol. Atmos. Phys.\/},  under review (arXiv:1303.2133).

\bibitem[Buizza  {\em et al.\/}, 1993]{btmp} Buizza, R., Tribbia, J.,
  Molteni, F. and  Palmer, T. (1993) Computation of optimal unstable
  structures for a numerical weather prediction system.  {\em Tellus
    A} {\bf 45}, 388--407.

\bibitem[Descamps {\em et al.\/}, 2009]{dljn} Descamps, L., Labadie,
C., Joly, A. and Nicolau, J. (2009) Ensemble Prediction at M\'et\'eo France
(poster introduction by Olivier Riviere) {\em 31st EWGLAM and 16th SRNWP
meetings,\/} September 28 --  October 1, 2009. Available at:
http://srnwp.met.hu/Annual\_Meetings/2009/download/sept29/morning/posterpearp.pdf   

\bibitem[Eckel and Mass, 2005]{em05} Eckel, F. A. and Mass,
  C. F. (2005) Effective mesoscale, short-range ensemble forecasting. {\em
    Wea. Forecasting\/} {\bf 20}, 328--350. 

\bibitem[Fraley {\em et al.\/}, 2010]{frg} Fraley, C., Raftery,
  A. E. and Gneiting, T. (2010) Calibrating multimodel forecast
  ensembles with exchangeable and missing members using Bayesian model
  averaging. {\em Mon. Wea. Rev.\/} {\bf 138}, 190--202.  

\bibitem[Fraley {\em et al.\/}, 2009]{frgs} Fraley, C., Raftery,
  A. E., Gneiting, T. and Sloughter, J. M. (2009) EnsembleBMA: An R
  package for probabilistic forecasting using ensembles and Bayesian model
  averaging.  {\em Technical Report\/} 516R, Department of Statistics,
  University of Washington.  Available at:
  www.stat.washington.edu/research/reports/2008/tr516.pdf 

\bibitem[Fraley {\em et al.\/}, 2011]{frgsb} Fraley, C., Raftery,
  A. E., Gneiting, T., Sloughter, J. M. and Berrocal, V. J. (2011)
  Probabilistic weather forecasting in R.  {\em The R Journal\/} {\bf
    3}, 55--63. 

\bibitem[Gebhardt {\em et al.\/}, 2011]{gtpb} Gebhardt, C., Theis,
  S. E., Paulat, M. and Bouall\`egue, Z. B. (2011) Uncertainties in
  COSMO-DE precipitation forecasts introduced by model perturbations
  and variation of lateral boundaries. {\em Atmos. Res.\/} {\bf 100},
  168--177.  

\bibitem[Gneiting, 2011]{gneiting11} Gneiting, T. (2011) Making and
  evaluating point forecasts.  {\em J. Amer. Statist. Assoc.\/} {\bf
    106}, 746--762.

\bibitem[Gneiting and Raftery, 2007]{grjasa} Gneiting, T. and Raftery,
  A. E. (2007) Strictly proper scoring rules, prediction and
  estimation. {\em J. Amer. Statist. Assoc.\/} {\bf 102}, 359--378.

\bibitem[Gneiting {\em et al.\/}, 2005]{grwg} Gneiting, T.,
  Raftery, A. E., Westveld, A. H. and Goldman, T. (2005) Calibrated
  probabilistic forecasting using ensemble model output statistics and
  minimum CRPS estimation. {\em Mon. Wea. Rev.\/} {\bf 133},
  1098--1118.

\bibitem[H\'agel, 2010]{hagel} H\'agel, E. (2010) The
  quasi-operational LAMEPS system of the Hungarian Meteorological
  Service. {\em Id\H oj\'ar\'as\/} {\bf 114}, 121--133.

\bibitem[Hor\'anyi {\em et al.\/}, 2006]{hkkr} Hor\'anyi, A,
  Kert\'esz, S., Kullmann, L. and Radn\'oti, G. (2006) The
  ARPEGE/ALADIN mesoscale numerical modeling system and its
  application at the Hungarian Meteorological Service.  {\em Id\H
    oj\'ar\'as\/} {\bf 110}, 203--227.  

\bibitem[Hor\'anyi {\em et al.\/}, 2011]{horanyi} Hor\'anyi, A., Mile,
  M. and Sz\H ucs, M. (2011) Latest developments around the ALADIN operational
  short-range ensemble prediction system in Hungary. {\em Tellus A} {\bf 63},
  642--651.

\bibitem[Justus {\em et al.\/}, 1978]{jhmg} Justus, C. G.,
  Hargraves, W. R., Mikhail, A. and Graber, D. (1978) Methods for
  estimating wind speed frequency distributions. {\em
    J. Appl. Meteor.\/} {\bf 17}, 350--353. 

\bibitem[Labadie {\em et al.\/}, 2012]{ldcm} Labadie, C., Descamps,L.,
  Cebron, P. and  Michel, Y. (2012) PEARP initialization with Ensemble
  Data Asssimilation and Singular Vectors. {\em International
    Conference on Ensemble Methods in Geophysical Sciences,\/}  
  Toulouse, France, November 12--16, 2012. Available at: 
http://www.meteo.fr/cic/meetings/2012/ensemble.conference/presentations/session07/\break
2.pdf

\bibitem[Leutbecher and Palmer, 2008]{lp} Leutbecher, M. and Palmer,
  T. N. (2012) Ensemble forecasting. {\em J. Comp. Phys.\/}  {\bf 227},
  3515--3539. 

\bibitem[Pinson and Hagedorn, 2012]{pinhag} Pinson, P. and Hagedorn,
  R. (2012) Verification of the ECMWF ensemble forecasts of wind speed
  against analyses and observations. {\em Meteorol. Appl.\/} {\bf 19},
  484--500.   

\bibitem[Raftery {\em et al.\/}, 2005]{rgbp} Raftery, A. E., Gneiting, T.,
  Balabdaoui, F. and Polakowski, M. (2005) Using Bayesian model
  averaging to calibrate forecast ensembles. {\em Mon. Wea. Rev.\/}
  {\bf 133}, 1155--1174. 

\bibitem[Sloughter {\em et al.\/}, 2010]{sgr10} Sloughter,
  J. M., Gneiting, T. and Raftery, A. E. (2010)  Probabilistic wind
  speed forecasting using ensembles and Bayesian model averaging. {\em
    J. Amer. Stat. Assoc.\/} {\bf 105}, 25--37.

\bibitem[Sloughter {\em et al.\/}, 2007]{srgf} Sloughter, J. M.,
  Raftery, A. E., Gneiting, T. and Fraley, C. (2007) Probabilistic
  quantitative precipitation forecasting using Bayesian model
  averaging. {\em Mon. Wea. Rev.\/} {\bf 135}, 3209--3220. 

\bibitem[Thorarinsdottir and Gneiting, 2010]{tg}  Thorarinsdottir,
  T. L. and Gneiting, T. (2010) Probabilistic forecasts of wind speed:
  ensemble model output statistics by using heteroscedastic censored
  regression. {\em J. Roy. Statist. Soc. Ser. A\/} {\bf 173},
  371--388.  

\bibitem[Toth and Kalnay, 1997]{tk} Toth, Z. and Kalnay, E. (1997)
  Ensemble forecasting at NCEP and the breeding method.  {\em
    Mon. Wea. Rev.\/} {\bf 125},  3297--3319. 

\bibitem[Wilks, 2011]{wilks} Wilks, D. S. (2011) {\em Statistical
    Methods in the Atmospheric Sciences.\/} 3rd ed., Elsevier,
  Amsterdam. 

\bibitem[Wilks and Hamil, 2007]{wh} Wilks, D. S. and Hamill,
  T. M. (2007) Comparison of ensemble-MOS methods using GFS
  reforecasts.  {\em Mon. Wea. Rev.\/} {\bf 135}, 2379--2390.
\end{thebibliography}
\end{document}